\documentstyle[preprint,aps,amssymb]{revtex}

\begin{document}

\title{ Nodes, Monopoles and Confinement\\
in $2+1$-Dimensional Gauge Theories}

\author{\bf {M. Asorey}, F. Falceto,
J. L. L\'opez and G. Luz\'on}
\address{Departamento de
F\'{\i}sica Te\'orica. Facultad de Ciencias\\
Universidad de Zaragoza.  50009 Zaragoza. Spain}

\maketitle
\begin{abstract}

	In the presence of  Chern-Simons interactions the wave functionals
of physical states in $2+1$-dimensional gauge theories vanish at a
number of nodal points. We show that
those nodes  are located at some classical configurations which carry
a non-trivial  magnetic charge. In abelian gauge theories this
fact explains why magnetic monopoles  are suppressed by
Chern-Simons interactions. In non-abelian theories it suggests a
relevant role for nodal gauge field configurations in the
confinement mechanism of Yang-Mills theories. We show that the
vacuum  nodes correspond to the chiral gauge orbits of
reducible gauge fields with  non-trivial magnetic monopole
components.
\end{abstract}
\hyphenation{systems}


The role of magnetic monopoles in the confinement mechanism  still
remains elusive  in QCD$_{3+1}$ in spite of the appealing conjectures
 based on the dual superconductor picture
\cite{th}.  However in 2+1 dimensions where they
play the role of instantons  their  contribution seems to be crucial
for confinement. In compact lattice QED$_{2+1}$ it has been  shown
that the logarithmic perturbative  Coulomb potential becomes linear
by means of Debye screening of electric charges in a monopole gas
\cite{poli} in a similar manner as vortices drive the
Kosterlitz-Thouless phase transition in the XY model \cite{bere}.

On the other hand, in 2+1 dimensional  gauge theories there exists the
possibility of having massive gluons while keeping gauge invariance.
This is possible because in 2+1 dimensional space-time massless and
massive vector-like particles have the same number of degrees of
freedom.  The generation of mass for gauge fields can be explicitly
achieved in local terms thanks to the peculiar properties of the
Chern-Simons term.

In such a case quarks are deconfined because no condensation
of pseudoparticles  can dramatically modify the exponential decay of
gauge propagators. In fact, it has been shown that  in compact
QED$_{2+1}$ the confinement of electric charges is traded by that of
magnetic monopoles \cite{Aff} (see also \cite{semen}),
and the magnetic superconductivity picture
of confinement is traded by a standard electric superconducting
scenario. In this sense a topological mass perturbation realizes an
electromagnetic duality  transformation.

In this note we elaborate on the role of magnetic monopoles in a
topologically massive Yang-Mills theory (YM$_{\rm TM}$). The
Hamiltonian of the theory in Schr\"odinger  representation reads
\begin{eqnarray}
H&=&-{\Lambda\over 2}\int {d^2x\over\sqrt{h}}\ {\rm Tr}\
\left|{\delta\over \delta A_\mu(x)}+ {ik\over
4\pi}\epsilon^{\mu\nu}A_\nu(x)\right|^2\cr&&
+{1\over 4\Lambda}\int d^2x\
\sqrt{h}{\rm Tr}\   F_{\mu\nu} F^{\mu\nu}(x)\label{eham}
\end{eqnarray}
where $h$ is the metric of the $2$D space.
Physical states are constrained by the Gauss law condition
\begin{eqnarray}
{D_\mu{\delta\over \delta A_\mu(x)}\psi(A)=
{ik\over 4\pi}\epsilon^{\mu\nu}\partial_\mu A_\nu(x)\psi(A)}\label{egauss}
\end{eqnarray}
where $D_\mu$ stands for the covariant derivative.

If $k$ is not an integer there are no solutions of Gauss law (\ref{egauss}).
This is the way the quantization of Chern-Simons coupling constant
$k$ appears in the canonical formalism  \cite{am}.
When $k$ is an integer
the solutions of the constraint equation (\ref{egauss})\ can be
geometrically characterized as sections of a non-trivial line
bundle defined over the space of gauge orbits
${\cal M}={\cal A}/{\cal G}$, i.e.
the space of gauge fields ${\cal A}$ moded out by the group of gauge
transformations ${\cal G}$ \cite{am}. The Chern class of this bundle
where the
physical states live ${\cal E}({\cal M},{\Bbb  C})$
is $c_1({\cal E})=k$. Now, because
the bundle is non-trivial any section (i.e. any physical state) must
vanish at some gauge field configurations \cite{kar}.
Such a behavior is in contrast with Feynman's claim on
the absence  of nodes in the ground state of pure  Yang-Mills theory
in 2+1 dimensions. He used that property, suggested by a formal
maximum principle, to argue that quarks are confined in 2+1
dimensions \cite{feyn}.

If there is any relationship between the absence of confinement
and the existence of nodes in  the vacuum state of  the theory with
a Chern-Simons term it is possible to establish a connection between
the configurations where the vacuum vanishes and permanent
confinement.

In quantum mechanics the existence of nodes in the ground state is
usually related to its degeneracy. In such a case, the {
position} of nodal points is not relevant because they change
from one state to  another.  However, in topologically massive gauge
theories  it has been claimed that the vacuum is not degenerated
\cite{car}. In
such a case, the nodal points are also unique and the corresponding
gauge configurations  would play a role in the  confinement
mechanism.

The relevance of  nodes  was anticipated in ref. \cite{kar}, however,
nodal configurations of the vacuum functional where unknown for a
long time. In the present note we provide the solution for this
longstanding problem.

In the abelian case, $G=U(1)$, there is no topological reason for
physical states to have nodes. If the space is compactified
to become a 2D sphere $S^2$ the first and second homotopy groups of
the orbit space vanish, $\pi_1({\cal M})=\pi_2({\cal M})=0$ \cite{ab}
\cite{amdos}  and, thus,  any line bundle over ${\cal M}$ is trivial.
Physical states are sections on a trivial bundle and, thus, they
can  be non-null for any gauge field configuration.

The space of orbits splits into  several disjoint pieces
($\pi_0({\cal M})={\Bbb  Z}$) each one containing abelian gauge fields
carrying  the same magnetic charge. Since the magnetic charge is
quantized by Dirac condition,  the different connected components of
${\cal M}$ are parametrized by an integer number $n$,
${{\cal M}=\cup_{n=-\infty}^\infty {\cal M}_n.}
$
{}From a topological viewpoint all the connected components ${\cal M}_n$ of
${\cal M}$ are equal. Actually,
they are diffeomophic to the component without
magnetic charge ${\cal M}_0$.

If all the sections of the bundle ${\cal M}\times {\Bbb  C}$ were physical
states, the Hilbert space would be a sum
${\cal H}=\oplus_{n=-\infty}^\infty {\cal H}_n$ of Hilbert spaces,  each one
corresponding to different monopole backgrounds. The energies from
each $n$-monopole sector would be shifted by $n^2/2\Lambda$ by the
effect of the potential term of the Hamiltonian.

However this is not the case because Gauss law imposes a very
restrictive condition on physical states. In fact, if we integrate
both sides of the Gauss law (\ref{egauss})\ we get
\begin{eqnarray}
{\int d^2
x\, {\partial_\mu{\delta\over \delta A_\mu(x)}\psi(A)=
{i k\over
4\pi}\int d^2 x \,B(A) \psi(A)},}
\label{abgl}\end{eqnarray}
where $B(A)=F_{12}(A)$ is the magnetic field strength.
The left hand side vanishes because under the integral we have a pure
differential whereas the  right hand side reduces to
$(kn/2)\psi(A)$,  $n$ being the magnetic charge carried out by
the gauge field $A$. Consequently, for gauge fields with non-trivial
magnetic charge ($n\neq 0$) the wave functional $\psi(A)$ must
vanish. This means that when $ k\neq 0$ only the sections over the
${\cal M}_0$ sector do correspond to physical states, i.e. ${\cal H}_{\rm
phys}\equiv {\cal H}_0$. The theory is exactly solvable and the vacuum
state reads
\begin{eqnarray}
\nonumber
\Psi_0(A) &= \exp&\Big\{
 \int\! d^2x \, {\rm Tr} \Big[{ik\over 4\pi}
 \partial^\mu A_\mu \Delta^{-1}B(A)
 \cr&&- {1\over 2\Lambda\sqrt{h}}
\,B(A)\Delta^{-1}\left(m^2+{\Delta}\right)^{_1\over^2}
B(A)\Big]\Big\}
\end{eqnarray}
for gauge fields without magnetic charge $\Phi(A)=\int d^2x
B(A) = 0$ and vanishes $ \Psi_0(A)=0$  for magnetic monopole
configurations ($\Phi(A)\neq 0$).
The spectrum corresponds to a free massive photon with mass $m=
k\Lambda/2\pi$ \cite{jack}.
The basic property involved in the above argument is that
constant gauge transformations do not  transform   abelian gauge
fields which implies the vanishing  of the left hand side of the
Gauss law
 whereas space constant temporal component of  gauge fields
($A_0$=cte) does  couple to the other components by means of both,
Yang-Mills and Chern-Simons, terms of the action, which  leads to
the equality of both sides of the Gauss law equation. In physical
terms  what happens is that the Chern-Simons term generates a
transmutation of magnetic charge into electric charge  which is
reflected in the anomalous terms of Gauss law.

The same physical argument applies for higher genus ($g>0$)
compactifications $\Sigma$ of the physical space.
In such  a case the topology of the space of gauge orbits without
magnetic charge, ${\cal M}_0$,  becomes more sophisticate and in fact for
$k\ne 0$ the bundle of physical states ${\cal E}({\cal M},{\Bbb  C})$ is
non-trivial. The quantization of the theory implies the existence of
additional nodes also on the orbit space ${\cal M}_0$. However in these
cases the vacuum is degenerated and the  configurations with
vanishing amplitudes $\psi(A)=0$ are not physically relevant. {}From
the above results, some of them anticipated  in Ref.
\cite{car}, we conclude  that  in abelian 2+1 dimensional gauge theories
Chern-Simons interactions are absolutely  incompatible with
monopoles.

In the case $k=0$ the theory reduces to a  pure Maxwell continuum
(``non compact'')  theory. External charges are confined by a
logarithmic potential.  Monopoles are not confined
and in fact they have a finite mass  $M=1/(2\Lambda)$. When photons
become massive by the effect of the Chern-Simons  interaction
electric charges are deconfined whereas magnetic monopoles decouple
from the physical degrees of freedom, i.e. their mass becomes
infinite and their correlators vanish. In this theory the dual
superconductor picture,  i. e. confinement/condensation of
magnetic/electric charges, is explicitly realized and the insertion
of the Chern-Simons term makes the transition from one regime to
another.

The  same analysis holds in the non-abelian theory
for gauge field configurations with non-zero
total magnetic charge.
However we will see that even in the sector of zero
net charge there are some configurations where
the states vanish.

Let us  restrict ourselves, for simplicity,
to the $SU(N)$ case, although our analysis can be
easily generalized for
arbitrary  gauge groups. Gauge fields are defined
on a trivial
bundle $P=\Sigma\times SU(N)$.
Reducible gauge fields $A$ can, actually, be defined
on  subbundles $P_r\left(\Sigma, U(N_1)\times\cdots\times U(N_r)
\right)
$ of $P$ with
structure group $ U(N_1)\times\cdots\times U(N_r)$. They  are
decomposed into a sum $A= A_1+A_2+\cdots+A_r$  of elementary gauge
fields $A_i$  with values in $u(N_i)$. The gauge field elementary
components $A_i$ of $A$ are defined on principal
bundles $P_i(\Sigma, U(N_i))$ whose first chern
classes $c_1(P_i)$ represent the magnetic charges
of the different  components of $A$.
Since $A$ is a connection with gauge group $SU(N)$
the total magnetic charge $\sum_{i=1}^r c_1(P_i)=0$ vanishes.

Reducible gauge fields are  invariant under the following group of
gauge transformations
\begin{eqnarray}
\label{mat}
\Phi_t=\pmatrix{{\rm e}^{i\mu_1t}I_1& 0& \cdots&0\cr
0& {\rm e}^{i\mu_2t}I_2& \cdots&0\cr
\cdots&\cdots&\cdots&\cdots\cr
0&0 & \cdots &{\rm e}^{i\mu_r t}I_r},
\end{eqnarray}
with $\mu_i=c_1(P_i)/N_i$ and where $I_i$ denotes the identity
matrix of $U(N_i)$.
Thus, the infinitesimal generator of the group $K_t$ of gauge
transformations $\phi( A_{\bar{ }})=\dot{\Phi}_t|_{t=0}$
satisfies  that $d_{A_{{}}}\phi( A_{})=0$,
and
\begin{eqnarray}
\label{cssl}{\int d^2x{\rm Tr}\, \phi( A)
D^\mu{\delta\over \delta A_{\mu} } \psi(
A_{})= 0, }
\end{eqnarray}
for any functional $\psi(A)$. In particular, for physical
states Gauss' law (\ref{egauss})\ implies that
  $\int {\rm Tr} \, \phi(
A_{ }) d A_{} \, \psi( A_{ })$ must vanish.
Now,
\begin{eqnarray}
\label{fin}\int {\rm Tr}\,  \phi( A_{})
d A &=&\int {\rm Tr} \, \phi( A)
 (d A+d_{A}A) \cr
&=&2\int{\rm Tr}\,  \phi( A_{ })   F(A)=4\pi
\sum^r_{i=1}{c_1(P_i)^2\over N_i}.
\end{eqnarray}
and, therefore, if one of the magnetic charges $c_1(P_i)$ of the
components of $A$ is
non null every physical state must vanish at that
gauge field configuration, i.e. $\psi(A)
=0$.

However, reducible configurations do not
exhaust all nodal configurations. The reason is that the topological
arguments leading to  the existence of nodes discussed at the
beginning also apply to the orbit space of {\it irreducible}
connections. Therefore, there should exist other  genuine
non-abelian configurations where physical states vanish. To find
those configurations a more elaborate dynamical argument is
required. In  general, nodal configurations will not be the
same for all physical states, only those with net magnetic charge
satisfy this property. Non-abelian, irreducible nodes depend
on the physical state we consider. In the following, we will
analyse the nodes of the vacuum state.

To study the vacuum state we need to minimize expectation value
the quantum hamiltonian
(\ref{eham}). For that it will be usefull to introduce
the chiral components of the connection.
Having fixed a 2-dimensional metric $h$ our 2D space $\Sigma$
acquires a complex structure and
this induces a chiral decomposition
$A=A_zdz+A_{\bar z}d\bar z$ of the
gauge field.
The component $A_{\bar z}$ defines an
holomorphic structure on the vector
bundle $E(\Sigma,{\Bbb  C}^N)$ associated to $P$.
Conversely,
once one fix an hermitian structure on $E$
any holomorphic structure on $E(\Sigma,{\Bbb  C}^N)$
defines a unique unitary connection $A$ on P.
This
correspondence  induces an isomorphism between the space of gauge
fields ${\cal A}$ and the space ${\cal A}_{\bar z}$ of holomorphic
structures on $E(\Sigma,{\Bbb  C}^N)$.

In terms of the chiral components of the gauge field
 the physical states that minimize the ``kinetic'' term of
the Hamiltonian (\ref{eham})\ are  of the form
\begin{eqnarray}
\label{holo}{
\psi(A)={\rm exp}\{ {ki\over 8\pi}\int dz\, d\bar{z}\,{\rm Tr}
A_{\bar{z}} A_z\} \xi( A_{\bar{z}}), }
\end{eqnarray}
$\xi(A_{\bar{z}})$ being any holomorphic functional of
$ A_{\bar{z}}$. The
restriction of the Gauss law to those states reads
\begin{eqnarray}
\label{css}{D_{\bar{z}}{\delta\over \delta A_{\bar{z}} } \xi(
A_{\bar{z}})= {k\over \pi}\partial_z A_{\bar{z}}\, \xi( A_{\bar{z}})
,}
\end{eqnarray}
and it is analogue to the Gauss law of pure Chern-Simons topological
field theory. In fact, one can identify these
states with those of the Chern-Simons
theory in holomorphic quantization \cite{gw}, \cite{car}.

In ${\cal A}_{{\bar z}}$  there is an action of a larger group of symmetries,
the group of chiral or {\it complex}
gauge transformations ${\cal G}^{\Bbb  C}$. The
action of $h\in {\cal G}^{\Bbb  C}$ on  $A_{\bar{z}}$
is given by
\begin{eqnarray}
\label{roto}{^h A_{\bar{z}}=h A_{\bar{z}}h^{-1}+
ih{\partial_{\bar z}}h^{-1},}
\end{eqnarray}
and the isomorphism between ${\cal A}$ and  ${\cal A}_{{\bar z}}$
induces an action of ${\cal G}^{\Bbb  C}$ in ${\cal A}$ that
extends the ordinary
or unitary gauge transformations in ${\cal G}$.

The relevance of these transformations
comes from the fact that integration of the Gauss
law (\ref{css})\ determines how
the states with minimal kinetic energy
change under chiral gauge transformations
of the gauge field.
In \cite{gw} it is shown that the states are
multiplied by a non-null factor depending
on $A$ and $h\in{\cal G}^{\Bbb  C}$.
Now, we have shown that physical states must
vanish for  reducible gauge fields with  magnetic monopole
components, therefore the  states with minimal kinetic energy also
vanish along their complex orbits.
Generically, the  gauge fields in
those orbits are  non-reducible
and thus we obtain this way a larger set
of nodes for these states.

One could argue that most of  the fields
belong to orbits of gauge fields without monopole components
which define
a dense set of ${\cal A}$. {}From a quantum point of view it means that
they are the most relevant configurations for the dynamical behavior
of the theory. However, it turns out that they are
the other orbits which are relevant for the discussion of
the structure of vacuum states.
In some sense, one can think of
the space of gauge fields expanded by complex gauge transformations
from configurations with non-trivial  monopole components as a
boundary  of the space of all gauge field configurations.
In this picture,
the topological effects would arise as boundary conditions to
be satisfied by the quantum states at those special configurations.

Atiyah and Bott \cite{ab} have studied
in detail the action of
the complex gauge group and they have
shown that the
chiral gauge orbits can be organized in strata
of ${\cal A}$.
The main stratum,
${\cal A}_0$, is made up of the gauge fields such that all
subbundles of the associated holomorphic
bundle $E$ have non-positive first Chern
class;  It is an open  dense submanifold
of ${\cal A}$. All  flat connections belong to this
stratum and the union of their complex gauge orbits
is dense in ${\cal A}_0$.
Then the
physical states with minimal kinetic energy (\ref{holo})\
are completely determined,
like Chern-Simons  states, by their values at flat connections
\cite{fg}.

In the case of the
sphere $\Sigma=S^2$ the complex gauge orbit of  the trivial connection
$A=0$ expands the main stratum ${\cal A}_0$ of ${\cal A}$.
Then there is a unique
minimal state   as in
Chern-Simons theory. Its  wave functional $\psi_0$ is  completely
determined by its value at $A=0$ and it vanishes nowhere
in ${\cal A}_0$. All other
orbits have a gauge configuration decomposable  into a sum
of monopole/anti-monopole gauge field components. Therefore,
the  physical states with minimal kinetic energy must vanish along
all complex gauge orbits unless ${\cal A}_0$.
This result is compatible with the fact
that for any value of
$\psi_0(0)$ the functional $\psi_0(A)$ converges to $0$ as $A$
approaches an orbit of a field configuration with monopole
components. A result that can be obtained following the
techniques of Ref. \cite{gw}.

For higher genus Riemann surfaces the structure
of the moduli space of flat connections is non-trivial and
the states with minimal kinetic energy are not uniquely determined
from the Gauss law (see \cite{fg} for the toroidal
topology).  There is a
finite-dimensional space of physical states and they can have
additional nodes for some particular flat connections.
This nodes, however, are not physically relevant as they
change from one state to another.

So far we have considered only the kinetic term
of the hamiltonian (\ref{eham})\ but the vacuum
state should minimize the whole hamiltonian,
which also includes a
potential term. To understand how this
term could affect the vacuum structure we need a more
elaborate argument.
The crucial remark is that, as we have seen,
the vacuum as any other
physical state has to vanish at reducible
gauge field configurations with monopole components,
but among those
configurations there are the
absolute minima
of the potential term when we restrict to the corresponding
strata \cite{ab}.
This can be understood from the
following remarks. The flow defined by the gradient field of the
potential term is tangent to the complex gauge orbits and such a
flow has critical points at
reducible configurations that are
solutions of Yang-Mills equations.
These critical gauge configurations can be found in any  strata
of ${\cal A}$, and
it can be shown that the negative modes of the
second variation of the potential at these critical points are
orthogonal to their strata, which implies that they are local
minima of the potential restricted to the those strata. The fact
that they are global minima follows from  the
results of Atiyah and Bott
\cite{ab}.

Vanishing of the wave functional $\Psi_0(A)$
along the chiral gauge orbits containing
solutions of Yang-Mills equation
with magnetic monopole components is,
then, necesary to minimize the expectation value of $H$. It is
not only required for the minimization of the kinetic term, but also
for that of the Yang-Mills potential term, and it is a consequence of
Ritz variational principle.
Both terms of the Hamiltonian, the kinetic and potential terms
conspire to make the vacuum to vanish on the orbits of gauge field
configurations with monopole components. It is obvious that we
can not extend this argument to higher energy states.

This result explains why in the limit of infinite topological
mass $\Lambda \rightarrow \infty$ we recover the Chern-Simons
states which by the same argument also vanish for the same
configurations, and are completely determined by their values
at flat gauge field configurations \cite{gw}.
A similar result holds for an arbitrary gauge group $G$.

In summary,
magnetic monopoles in YM$_{TM}$ are
suppressed in any physical state by kinematical constraints,
but the gauge field
configurations on their complex gauge orbits
are also suppressed in the
vacuum state. They only give non-trivial contributions to excited
states. Since the YM$_{TM}$ is not confining it is
natural to speculate about the
connection between the existence of nodes and
the absence of confinement induced by the
Chern-Simons interaction. If this connection exists
an important role will be played by those configurations in
the confinement mechanism for pure gauge theories. So far, most of
the confinement scenarios gave a leading role  to
magnetic monopoles. {}From the above analysis it might
be inferred that the gauge fields which are chiral
gauge equivalent to those monopoles
also play a  relevant role. This opens a new possibility for
understanding the mechanism of permanent confinement in 2+1
dimensional  gauge theories.

\acknowledgements

M.A. thanks  Prof. M. Atiyah for discussions on  the
possible relevance of reducible connections.
 J. L. L. was supported by a MEC fellowship  (FPI
program)  and G.L. by a CONAI (DGA) fellowship. We also acknowledge
to   CICyT
for partial financial support under grants AEN93-0219 and AEN94-0218.


\end{document}